\documentclass[%
 aps,
]{revtex4-1}

\usepackage{graphicx}
\usepackage{dcolumn}
\usepackage{bm}

\begin{document}


\title{Improved depth resolution and depth-of-field in temporal integral imaging systems through non-uniform and curved time-lens array}

\author{Farshid Shateri}

\author{Shiva Behzadfar}
\author{Zahra Kavehvash}
 \affiliation{Department of Electrical Engineering, Sharif University of Technology, Tehran, Iran.}

\date{\today}

\begin{abstract}
Observing and studying the evolution of rare non-repetitive natural phenomena such as optical rogue waves or dynamic chemical processes in living cells is a crucial necessity for developing science and technologies relating to them. One indispensable technique for investigating these fast evolutions is temporal imaging systems. However, just as conventional spatial imaging systems are incapable of capturing depth information of a three-dimensional scene, typical temporal imaging systems also lack this ability to retrieve depth information\textemdash different dispersions in a complex pulse. Therefore, enabling temporal imaging systems to provide these information with great detail would add a new facet to the analysis of ultrafast pulses. In this paper, after discussing how spatial three-dimensional integral imaging could be generalized to the time domain, two distinct methods have been proposed in order to compensate for its shortcomings such as relatively low depth resolution and limited depth-of-field. The first method utilizes a curved time-lens array instead of a flat one, which leads to an improved viewing zone and depth resolution, simultaneously. The second one which widens the depth-of-field is based on the non-uniformity of focal lengths of time-lenses in the time-lens array. It has been shown that compared with conventional setup for temporal integral imaging, depth resolution, i.e. dispersion resolvability, and depth-of-field, i.e. the range of resolvable dispersions, have been improved by a factor of 2.5 and 1.87, respectively.
\end{abstract}
\maketitle
\section{Introduction}
\begin{figure*}
  \centering
  \includegraphics[width=\linewidth]{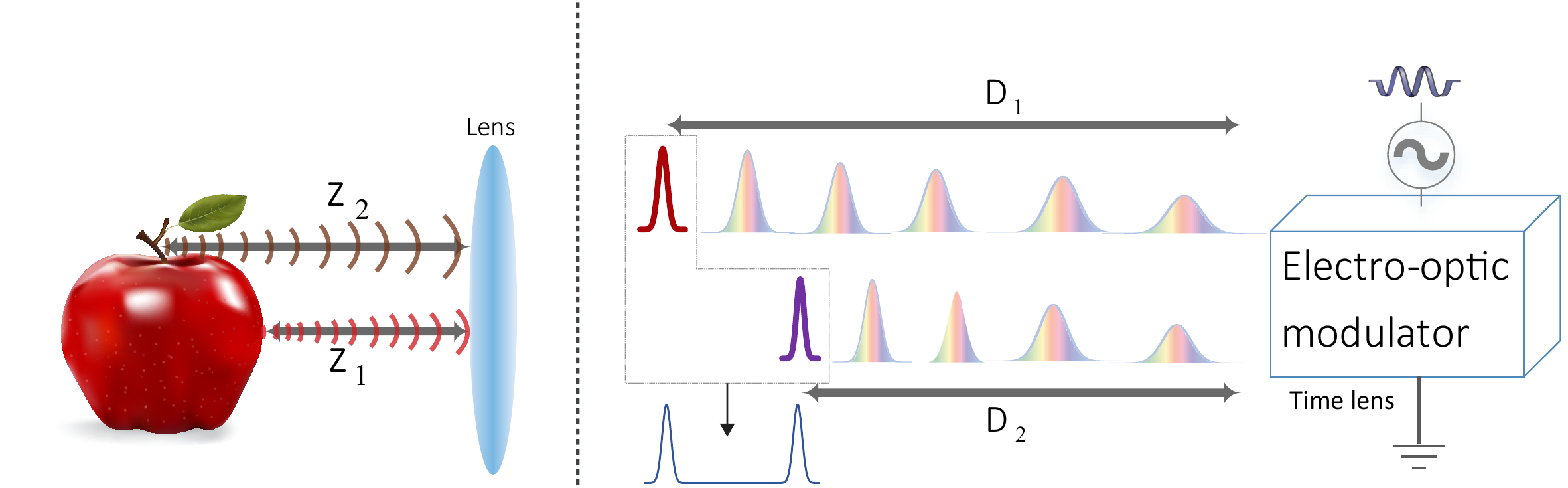}
  \caption{Visualization of depth concept in spatial (left) and temporal (right) optics. In left, two different depth distances of $z_1$ and $z_2$ of a same object are depicted and in right, two different GDD values ($D_1$ and $D_2$) of a signal (blue curve) are illustrated.}
  \label{fig1}
\end{figure*}

Temporal imaging systems (TIS), since the last two decades, have gained variant applications ranging from optical communications \cite{comm}, optical data acquisition and compression \cite{comp,stretch} to ultrafast microscopy \cite{microb}, quantum information processing \cite{q1,q2}, spectroscopy \cite{spec1,spec2}, and etc. \cite{kolner1,kolner2,kolner3,salem}. Apart from numerous studies trying to expand and develop the applications of TISs, researchers have also focused on their quality and performance improvement in view of resolution enhancement \cite{res1,res2} and aberration reduction \cite{abe}, among others.\\
Yet, one important quality feature of a TIS, which has remained to be improved, is to make them able to capture information laid in various depths. By depth, here, it means different dispersion lengths of each pulse or the group delay dispersion (GDD) that these pulses experience until they reach to the TL. These pulses can be viewed as different fragments of a bigger pulse. Therefore, in a TIS, just as a 3-dimensional (3D) spatial object is composed of varying diffraction lengths from the lens, a multi-dispersion pulse (or in some sense, a 3D pulse) consists of different dispersion lengths or GDDs (this concept has been visualized in Fig. \ref{fig1}). Thus, temporal depth imaging systems (TDIS) are needed to acquire this information.\\
TDISs have applications in better studying and investigating the evolution of fast and rare phenomena. Multi-pulse formation phenomenon which happens in passive mode-locked lasers, and results in noise-like pulses (NLP), is attributed to dispersion and saturable gain of spectral filters \cite{11}. Therefore, extracting the dispersion of each pulse in the NLP helps in better investigating the evolution of pulses in the cavity and can improve the future designs of ultrafast fiber lasers. Furthermore, rogue-waves \cite{rog1,rog2} are other extreme events that may occur in optical fibers as a combined effect of nonlinearity and dispersion. The same phenomenon may occur unpredictably in oceanic waves that may damage vessels. TDISs can provide us with a high temporal depth resolution that enables a more accurate investigation of rogue waves. This analysis includes extracting the dispersion of different random pulses that have been generated during the evolution of the rogue-waves.\\
In spatial optics, the well-known 3D imaging is a technique to retrieve the depth information of a full 3D scene. Among many 3D imaging methods, integral imaging (InI) is one of the most promising approaches due to it's ability to record full hologram of the 3D object with incoherent light through spatial lenses, without the need to record the phase of light (for a more in-depth analysis of spatial InI see Ref. \cite{depth1,depth2}). InI works based on the optical back propagation principle. For an object in a specific depth distance, the back projected elemental images (EI) through the same lens array (LA) as in the pick-up stage will overlap in the same depth distance resulting in a bright and clear image of that object. Still, reconstructing in other depth values results in non-overlapping images of each lens making the image of that object to become blurred in other depth distances. This will specify the depth of each object in the reconstructed 3D image.\\
Very recently, a research led by Fridman \textit{et. al.} has introduced and analyzed both theoretically and experimentally a technique to extract temporal depth information \cite{TDIS}. They used a method based on the relation between the lateral space of the images of two input pulses of different depth (dispersion) and their depth difference and lens location. The slope of the curve relating the separation between pulse images in each EI and the lens temporal location represented their depth (dispersion) difference. However, their method has been applied to a simple input signal including two pulses of different dispersion values while the slope of the curve yields this dispersion difference. For a more realistic signal which includes numerous dispersion values extracting the depths difference of each two pairs would be computationally and experimentally cumbersome.\\
In this paper, we approach this problem with the temporal counterpart of the spatial InI. More importantly, we have introduced two methods to improve the depth resolution and depth-of-field (DoF) of temporal InI. In spatial optics, many researchers have introduced various methods to overcome some shortcomings of spatial InI such as low resolution and limited depth-of-field region. Two of these methods which are highly effective are based on a curved lens array instead of a flat one and a non-uniform lens array in which different lenses have different focal lengths \cite{resub1,dof1,dof2}. We have adopted these solutions into temporal InI systems.\\
The rest of this paper is organized as follows: In section \ref{theo}, we briefly discuss pickup and reconstruction steps for both spatial and temporal InI systems. The proposed methods for enhancement of DoF and depth resolution are presented in Section \ref{noncurve}. In section \ref{result}, results from numerical simulations have been provided. Finally, in section \ref{con} the proposed methods and their results have been concluded.\\
\section{Spatial and temporal integral imaging: forward and reconstruction steps}\label{theo}
InI technique is composed of two steps. First, the forward imaging step in which EIs are obtained and second, the reconstruction step that uses the EIs captured by the first step to reproduce or reconstruct the full 3D scene. In this section, we will briefly discuss these steps of spatial InI systems and based on that, we will discuss the forward and reconstruction steps in temporal InI using the so-called space-time duality in electromagnetics. We will also show mathematically how depth information in spatial InI (diffraction information) and depth information in temporal InI (dispersion information) is laid in EIs. After showing how spatial InI could be generalized to the time domain in this section, two separate methods would be introduced to improve the performance of temporal InI in the next section.

\subsection{Forward and reconstruction model of a spatial integral imaging system}\label{spatial depth}
\begin{figure*}[!ht]
  \centering
  \includegraphics[width=\linewidth]{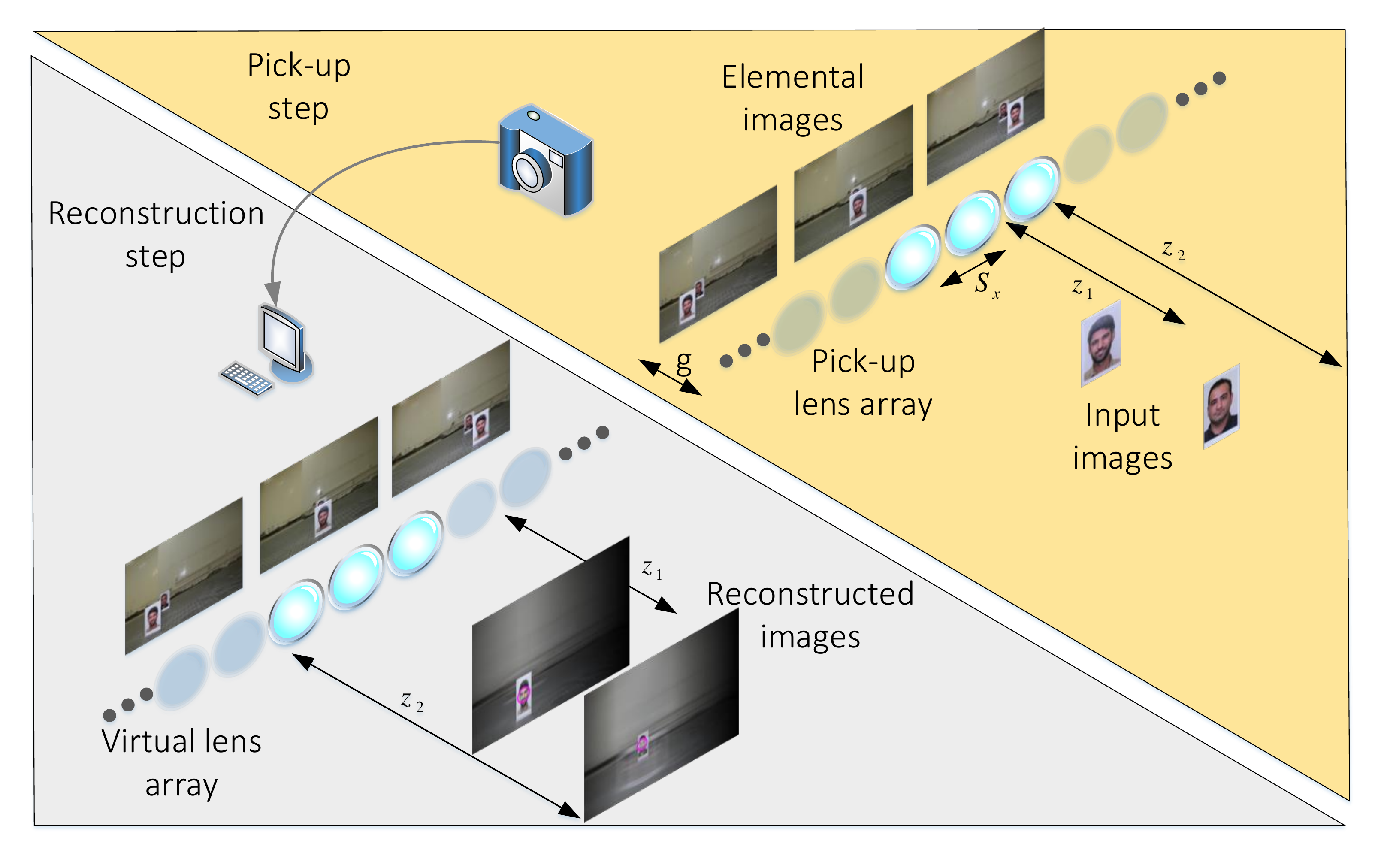}
  \caption{One-dimensional scheme of spatial InI. In the yellow rectangle (pick-up step), EIs are obtained through each lens and then these EIs are transferred to the reconstruction step (gray rectangle) in order to virtually reconstruct the image of a desired depth length. Virtual LA's specifications are the same as the specifications of the LA in the pick-up step. As can be seen in the gray rectangle, reconstruction at lengths equal to $z_1$ and $z_2$ leads to an accumulated intensity of the corresponding images at these lengths in the pick-up step (Images are obtained in our InI laboratory).}\label{fig2}
\end{figure*}

The main goal of an LA-based InI system is to make it possible to record and reconstruct depth information through capturing 2D EIs with incoherent light. As shown in Fig. \ref{fig2}, forward step of spatial depth imaging is based on utilizing an LA or a pinhole array. Each 2D slices of a 3D scene are mapped by this LA onto its image plane and therefore, we will have $m$ images, namely EIs, where $m$ is the number of lenses in the LA. The difference between EIs of the same 2D depth slice is the lateral position differences induced by the lenses in the array. Besides, EIs of two distinct 2D depth slices of a 3D scene differ by magnification factor since it depends on the distance between object to the lens and the distance from the lens to the image plane. Thus, we will have a bunch of EIs that, as will be discussed later in this section, depth information of the 3D scene is stored in them.
As will be discussed in the next subsection, in a similar manner, and by utilizing temporal ray optic notion called temporal rays, we will describe the forward and reconstruction model of a temporal InI system. Temporal ray representation is a useful tool to better understand the temporal optics concepts by visualizing it just like spatial geometrical optics. Given that each spectral component of a pulse travels at a particular velocity in a dispersive media, group delay in traveling wave coordinates for each spectral component has a direct relationship with group delay dispersion (this concept has been discussed pedagogically in Ref. \cite{abe}). Therefore, to each spectral component, we can attribute a line with a particular slope in the space-time coordinates.\\
Now, based on Ref. \cite{depth1}, we are first going to briefly explain the procedure to reconstruct the 3D spatial image from all the information stored in the EIs and then generalize the same procedure for a temporal InI system. By this computer synthesized reconstruction, we can achieve any voxel values at arbitrary depth distances. This computational reconstruction uses geometrical optics to explain the procedure of retrieving depth information from the EIs captured optically by the pickup LA. The procedure of reconstruction starts with computational (virtual) LA called display LA which has the same specifications as the pickup LA. This LA is feeded by the EI at the $p$'th row and $q$'th column ($I_{pq}$) and produces the inversely mapped image at distance $z$, namely $O_{pq}(x,y,z)$. According to some simple geometrical optics laws, $O_{pq}(x,y,z)$ can be expressed as \cite{depth1}:
\begin{equation}\label{o}
  O_{pq}(x,y,z)=\frac{I_{pq}({-\frac{x}{M}}+({1+\frac{1}{M}})s_x{p}, {-\frac{y}{M}}+({1+\frac{1}{M}})s_y{q})}{(z+g)^2+[(x-s_xp)^2+(y-s_yq)^2]({1+\frac{1}{M}})^2},
\end{equation}
where M is the magnification factor equal to $-g/z$, $I_{pq}$ is the pth and qth elemental image, $s_x$ and $s_y$ are the EI pitches along $x$ and $y$ directions and $g$ is the distance between the LA plane and the sensor.\\
Based on this relation, the location of the image of each $(x,y,z)$ point in the $p,q$ EI depends both on the depth of that point and the EI number. Therefore, for the points in a specific depth distance, the lateral location of their image on the EI array varies with a specific periodicity which is directly related to the depth value. The superposition of all the inversely mapped EIs yields the whole 3D image at all $(x,y,z)$ locations in 3D space as:
\begin{equation}\label{osum}
  O(x,y,z)=\sum_{p=0}^{m-1}\sum_{q=0}^{n-1}{O_{pq}(x,y,z)}.
\end{equation}
Fig. \ref{fig2} shows the computational reconstruction method on a display plane at two different depths. As depicted in this figure, each EI is inversely projected through each of the lenses in the virtual LA. The inverse projection of all the EIs through all lenses in the array at distance $z=z_1$ and $z=z_2$ leads to the formation of focused images of the objects located at the same depths while the superposition of the inversely mapped images of objects of other distances destructs each other. Therefore, we can obtain the full 3D scene by repeating the same process with computational virtual LA in all depth distances. Fig. \ref{fig2} shows the result of this procedure graphically. In this image which is from our previous work \cite{ghane}, we can observe that reconstruction at a particular distance leads to only one of the objects which is located at the same depth distance to become apparent.\\
One important point to note here is that every 2D image plane of a 3D scene experiences a different diffraction value up until it reaches to pickup LA. Now, by the fact that diffraction and dispersion are mathematically analogous with each other, in the next subsection, we introduce the temporal integral image reconstruction method and use it to obtain the depth information of a multi dispersion pulse.

\subsection{Forward and reconstruction model of a temporal integral imaging system}\label{temporal depth}
\begin{figure*}[!ht]
  \centering
  \includegraphics[width=\linewidth]{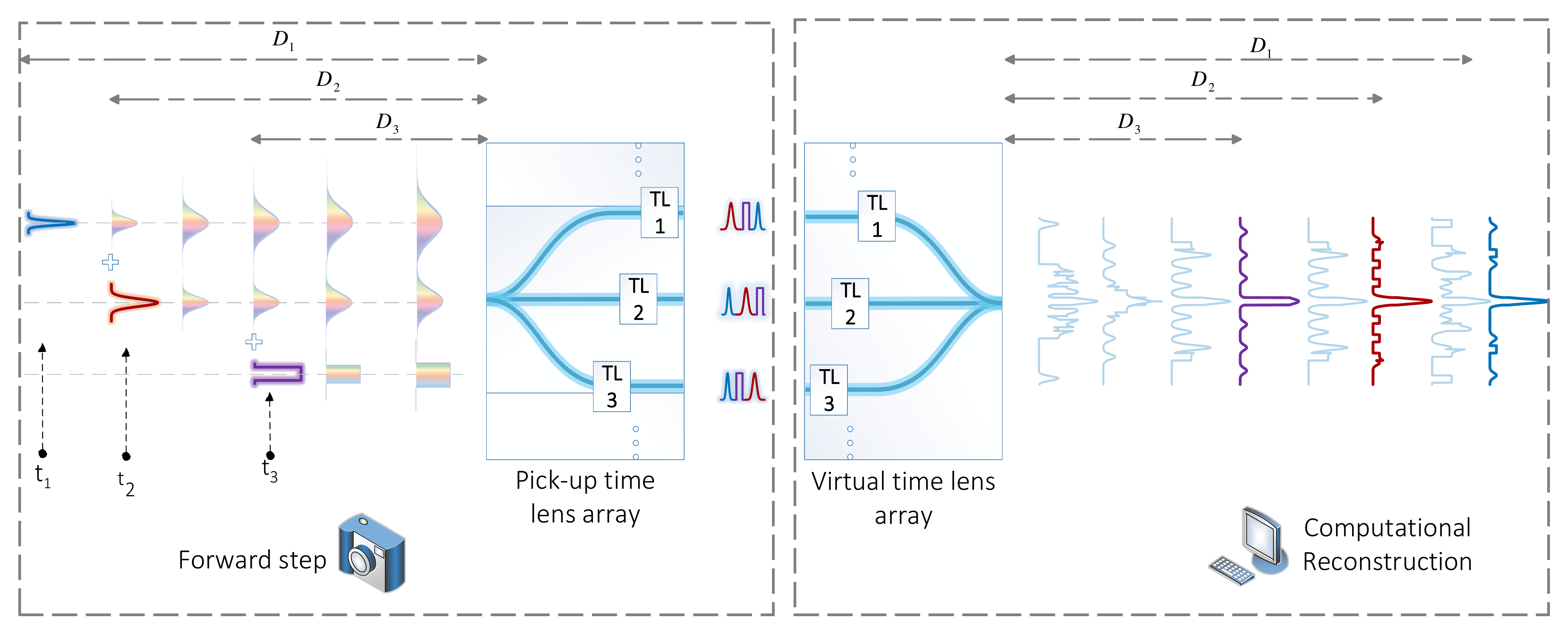}
  \caption{Schematics of temporal InI. In the pick-up step (left), the input signal consists of two Gaussian pulses (red and blue) and one rectangular pulse (purple) at dispersion depths of $D_1$, $D_2$, and $D_3$, respectively, which together form a multi-dispersion pulse. Then, temporal EIs are obtained through a TL array, and after passing the EIs through the virtual TL array (right), at dispersion lengths equal to $D_1$, $D_2$, and $D_3$, one can observe the purple, red, and blue curves which are well-correlated with their corresponding input pulses.}\label{fig3}
\end{figure*}

The information provided by spatial InI is the information that tell us what the image of a 3D scene is in a particular distance from the LA. In other words, given that this distance have a direct relationship with the amount of diffraction that the image experiences on its way to the LA, one can say that the role of InI system is to distinguish images that have experienced different values of diffraction.\\
By considering space-time duality, it is now reasonable to think of temporal InI as a technique that distinguishes pulses composed of sections with different dispersions. The main difference between spatial and temporal InI, however, is that the lateral dimension in spatial InI is 2D while for it's temporal counterpart it is 1D, i.e. time. Fig. \ref{fig3} shows an illustration of this method. As depicted in this figure, first, a single Gaussian pulse propagates along the way to the TL array and then at two points on its way to the TL array, two other pulses are added to it. Therefore, the complete pulse that travels to the TL array is composed of two Gaussian and one rectangular pulses that have experienced different dispersions (this is quite similar to the spatial case where an object consisting of three points in three different depth distances experience different diffraction values until the lens). Each TL is responsible for imparting a quadratic phase term to the input pulse just as a conventional spatial lens multiplies this phase to an input beam of light. These devices are realized via different methods among which the most conventional and commercially available ones are TLs produced by nonlinear optics methods such as four-wave-mixing (FWM). In Ref. \cite{salem}, these temporal optic apparatuses have been explained.\\
In spatial InI, the pickup LA is composed of lenses wherein each lens has a shifted quadratic phase profile due to its lateral position. For example, the central lens has a phase profile ${\phi}_{0,0}(x,y)\propto{x^2+y^2}$ while the phase profile of it's adjacent left or up lenses are ${\phi}_{-1,0}(x,y)\propto{(x+1)^2+y^2}$ and ${\phi}_{0,1}(x,y)\propto{x^2+(y-1)^2}$, respectively. By comparison, the phase profile of TLs in the TL array is, for example, ${\phi}_{0}(t)\propto{t^2}$ and ${\phi}_{1}(t)\propto{(t-1)^2}$, for the central TL and it's right adjacent. Note again that the lateral dimension in the temporal InI has been reduced from 2 dimensions in spatial imaging to 1 dimension. Similar to its spatial counterpart, the image captured through the $p$th TL in the temporal LA is called temporal EI, shown hereafter by $\tilde{I}_{{p}}$.\\
Now, the same procedure as spatial reconstruction could be performed for temporal depth reconstruction. As mentioned earlier in this section, the information that we deal with in the temporal InI is the depth information that stems from the particular dispersion values that different parts of a complex pulse have experienced until they arrive at the pickup TL array. Just as spatial case, the inversely mapped image of the ${p}$th EI can be written as:
\begin{equation}\label{otem}
  O^t_{{p}}(t,D)=\frac{\tilde{I}_{{p}}({-\frac{t}{\tilde{M}}}+({1+\frac{1}{\tilde{M}}})\tilde{s}_t{{p}})}{(D+{\tilde{g}})^2+(t-\tilde{s}_t{{p}})^2({1+\frac{1}{\tilde{M}}})^2},
\end{equation}
where in this case $\tilde{S}_t$ can be regarded as the temporal aperture of each TL in the array, $\tilde{g}$ is the dispersive distance between the TLs and the detector, and $\tilde{M}$ is the magnification factor which is equal to $-\tilde{g}/D$. By summing up all these inversely mapped images at a particular dispersion length, $D$, through the virtual LA, we have:
\begin{equation}\label{osumtem}
  O^t(t,D)=\sum_{p=0}^{m-1}{O^t_p(t,D)},
\end{equation}
which is the reconstructed pulse at an arbitrary $GDD=D$.
 The temporal reconstruction step is also illustrated in Fig. \ref{fig3}. The virtual LA here is made up of a virtual array of TLs which has the same specifications as the pickup temporal TL array. As depicted in Fig. \ref{fig3}, EIs are mapped onto a specific image plane of the virtual TL array corresponding to a specific dispersion value in a way that the amplitude of the desired pulse to be reconstructed (the pulse with that dispersion value) is built up and pulses with other dispersion lengths become less intense compared to the pulse under reconstruction.

\section{Using nonuniform and curved time lens array to improve DoF and depth resolution}\label{noncurve}
Despite being one of the most promising techniques in 3D imaging, InI systems have shortcomings in terms of depth resolution and DoF \cite{new1,new2,new3,new4}. In spatial optics, many researchers have offered solutions to these drawbacks. Two of the most prominent approaches are using a curved LA and a nonuniform LA in which lenslets have different focal lengths \cite{new5,new6}. In this section, we attempt to generalize these solutions to the case of temporal InI. In the next section, the enhancement of DoF and depth resolution via these two methods is proved through numerical simulations.

\subsection{Depth-of-field enhancement via non-uniform time-lens array}\label{non}
\begin{figure}[!ht]
  \centering
  \includegraphics[width=8cm]{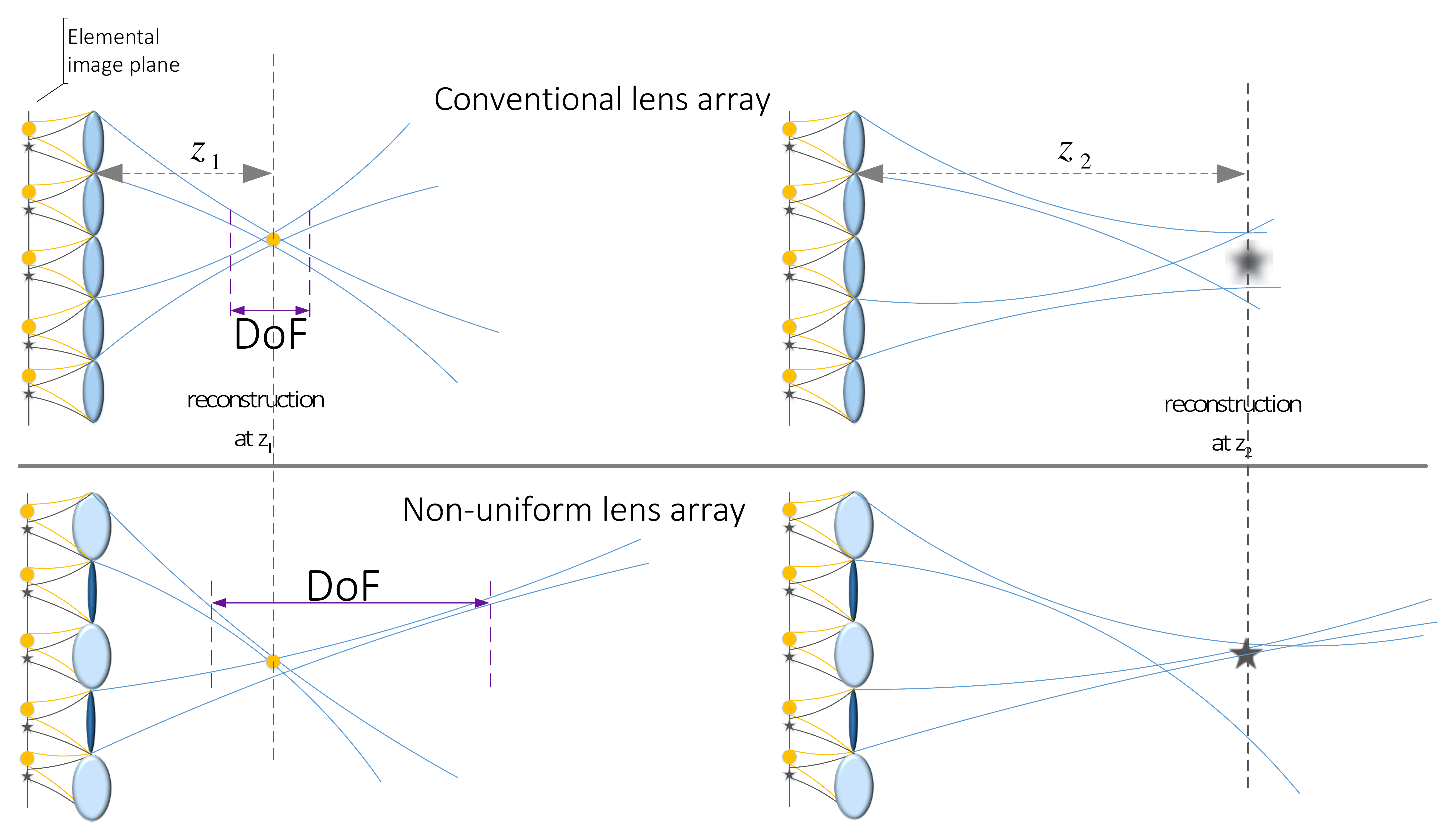}
  \caption{Schematic representation of DoF enhancement using a non-uniform lens array. Top figure shows a conventional InI system in which reconstruction out of DoF region has led to a blurred image (upper gray star) and bottom figure shows improvement in DoF region of the system, in turn, has led to a high resolution of the reconstructed gray star image.}\label{nonuniform}
\end{figure}

A single lens has a limited DoF, an area wherein the image is highly focused. In the pick-up step of InI, when the objects have a relatively large distance with respect to each other (in both temporal and spatial InI), it is probable that the whole 3D scene cannot be placed in the DoF region of each lens. Therefore, the reconstructed object would have a low resolution in some depth distances placed out of the DoF region. As mentioned earlier, a solution for this is to use an LA with different focal lengths. Since the lenses have different focal lengths, their DoF also varies. Thus, the overall DoF region of the LA which is the superposition of all the DoFs of single lenses becomes wider compared to DoF of an LA consisted of uniform lenses. As could be seen in Fig. \ref{nonuniform}, incorporating lenses with different focal lengths leads to a wider DoF. As could be seen in this figure, two signals of considerably different depth distances have been reconstructed with a high lateral resolution in the nonuniform set-up. In contrast, the conventional uniform lens array yields a low-resolution image of the star as its depth falls out of the DoF of the lenses. The effect of using non-uniform TL array in imaging a wider range of dispersion values with good resolution is demonstrated through numerical simulations in the next section. \\

\subsection{Depth resolution enhancement via a curved time lens array}\label{curv}

When the input object is complicated and made up of various depth features, obtaining those information with high depth resolution becomes challenging. One of the main difficulties in spatial InI systems (this is also true for temporal InI systems as will be discussed in this section) is the limited space of the sensor corresponding to each lens image (EI), which limits the depth resolution as will be discussed in the following.\\
In order to describe the concept of depth-resolution in a 3D spatial/temporal InI system and the effect of system parameters on its value, we proceed with the analysis performed in Ref. \cite{depth3}, which has approached the relationship between the 3D object and the EIs through the following steps. We start with the integral image of a point source object located at $(x_s,z_s)$ which is composed of a series of replicas at





\begin{equation}\label{j1}
x_m(z_s)=M_sx_s+mT_s
\end{equation}
spread in different EIs (here for simplicity the $y$ coordinate is neglected). In this relation, $M_s=-g/z_s$ is the magnification factor, $T_s$ is the pick-up period and is defined as the distance between the replicas of the images of the point source and depends on the LA's pitch ($S_x$) as
\begin{equation}\label{j2}
  T_s=(1+g/z_s){S_x}.
\end{equation}

Therefore, the integral image of the 1D object centered at $z_s$ can be calculated as:

\begin{equation}\label{j3}
  I_s(x)=rect(\frac{x-x_s}{\Delta{x}})\sum(\delta{(x-x_m)}\otimes{\frac{1}{M_s}}O(\frac{x}{M_s})),
\end{equation}

in which $\otimes$ indicates the convolution operator, $\Delta{x}=(n_x-1)T_s$ ($n_x$ is the number of lenses that contribute to the integral image) , and $O(x)$ is the object's intensity distribution.


Therefore, considering the above relation, having adequate number of EIs in hand, the periodicity shows itself clearly and thus the depth value could be retrieved.\\
As for spatial InI, to show how temporal depth information lies in the periodicity of integral image we introduce the temporal versions of Eq. \ref{j1}-\ref{j3}. Considering the integral image of a pulse located at $(t_s,D_s)$ (the lateral coordinate is now $t_s$ (time) and the depth here is the GDD denoted as $D_s$), it is composed of replicas at $\tilde{t}_m$ as:
\begin{equation}\label{jt1}
\tilde{t}_m(D_s)=\tilde{M}_st_s+m\tilde{T}_s,
\end{equation}

where $\tilde{M}_s=-{\tilde{g}}/{D_s}$ is the temporal magnification factor, ($\tilde{g}$ is the GDD from temporal LA to image plane ), and $\tilde{T}_s$ is the lateral distance (time) between these replicas which itself could be written as:

\begin{equation}\label{jt2}
 \tilde{T}_s=(1+{\tilde{g}}/D_s)S_t.
\end{equation}

Therefore, as in spatial case, the temporal integral image of a single dispersion pulse centered at temporal depth $D_s$ can be written as:

\begin{equation}\label{jt3}
  \tilde{I}_s(t)=rect(\frac{t-t_s}{\Delta{t}})\sum(\delta{(t-t_m)}\otimes{\frac{1}{\tilde{M}_s}}O(\frac{t}{\tilde{M}_s})),
\end{equation}
in which ${\Delta{t}}=(n_t-1)\tilde{T}_s$ and $n_t$ is the number of TLs that contribute to the temporal integral image. Again, we can infer from the last equation that depth information of a complex pulse composed of different dispersions appears itself in the periodicity of the integral image of each depth.\\
Based on the performed analysis, retrieving the information of each depth is possible through calculating the Fourier of the temporal integral image:
\begin{figure}
  \centering
  \includegraphics[width=8cm]{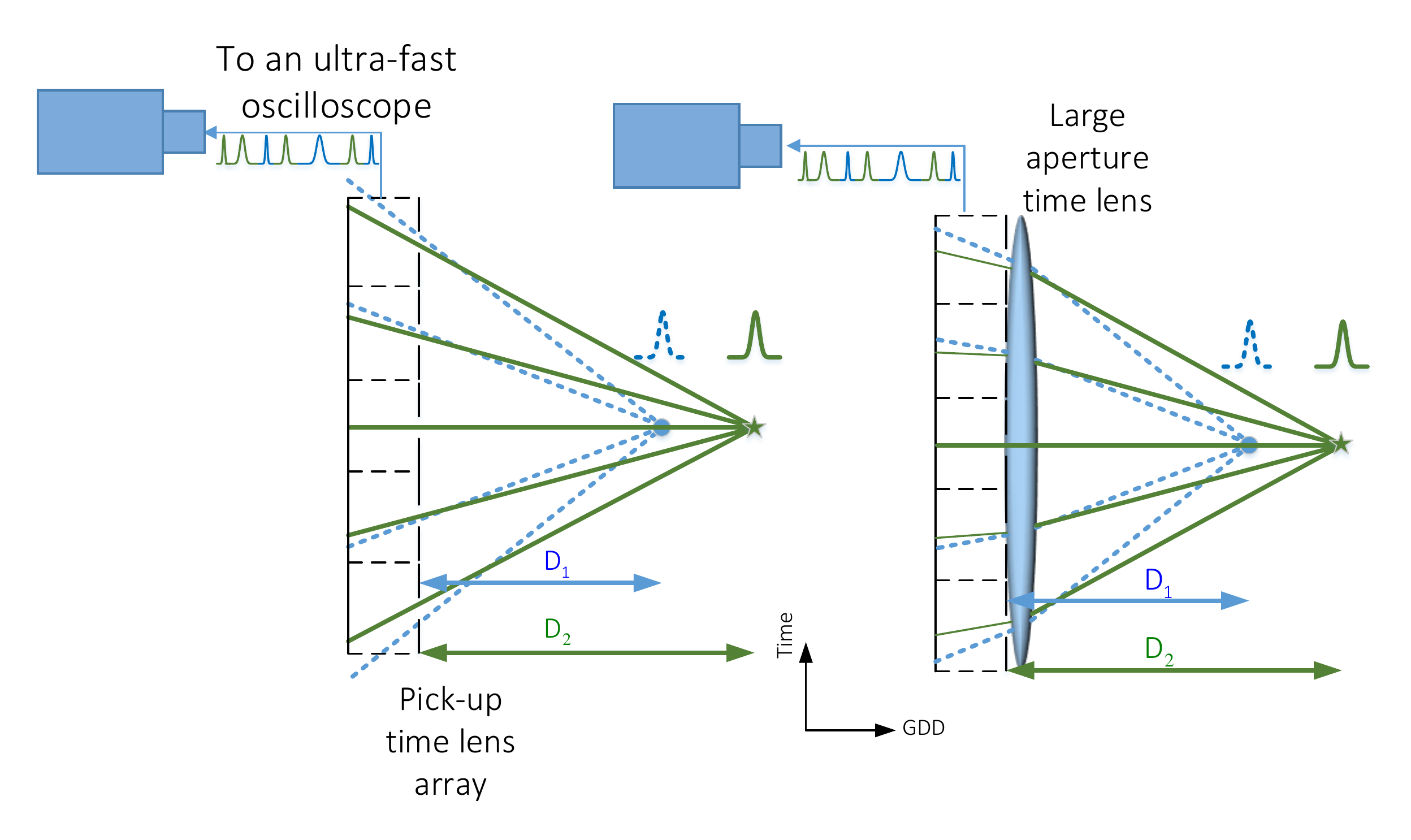}
  \caption{Comparison of a conventional and curved InI system. The curving effect is performed by a large aperture time-lens in front of the time-lens array as depicted above. Using this structure, temporal rays falling out of the recording region in the conventional InI system can now become captured in the corresponding temporal region of the ultrafast recorder.}\label{curve1}
\end{figure}

\begin{equation}\label{curv1}
  {\tilde{I}_s}(\omega)=\Delta{t}sinc(\Delta{t}\omega){\otimes}\text{exp}(i\frac{2{\pi}{\tilde{g}}}{t _s}\omega{t_s})\tilde{O}(\tilde{M}_s{\omega})\frac{1}{\tilde{T}_s}\sum(\omega-\frac{m}{\tilde{T}_s}),
\end{equation}
where $\otimes$ denotes the convolution operation. As could be noticed in this relation, each depth image shows itself as a specific frequency component of this spectrum which could be retrieved through Fourier filtering. Still, this is possible if enough number of periods of each depth image occurs in the integral image to show itself as a narrow-band and separable frequency component in $\tilde{I}_s(\omega)$. On the other hand, if the TL array is capable of capturing more rays from a depth slice of a pulse of specific dispersion value, the $rect$ function in Eq. \ref{j3} becomes larger in time and therefore the $sinc$ function in Eq. \ref{curv1} becomes narrower in the Fourier domain. Therefore, depth information of a specific depth distance could be easily filtered out from the spectrum of temporal integral image. Using this Fourier analysis on the temporal integral image, it is now apparent that capturing a higher number of periods of the image of each pulse of specific dispersion in the integral image leads to its better depth resolvability.\\
Nevertheless, in conventional InI systems, the size of each EI is limited due to preventing the overlap between neighboring EIs. This limited space for capturing the perceived rays by each lens leads to the loss of some says of high inclinations angles spatially those coming from nearer objects. This effect could be seen for some rays coming from the circular object in Fig. \ref{curve1} (a). This, in turn, limits the number of periods captured from its image in the integral image which as discussed above limits its depth resolvability. The same limitation exists in the temporal InI system but for a different reason. In the pick-up stage of a temporal InI, a pulse propagates in a number of branches until it passes through the TL in each branch. Then, all of the pulses in each branch reach to a single ultrafast detector. Therefore, the output of each TL has a limited time equal to the TL aperture to be recorded on the sensor. Other parts of the pulse outside this time span will be lost. These parts of the pulse again correspond to the temporal rays with high inclination angles which are probably coming from pulses of lower dispersion values, i.e., nearer depth distances. This effect again limits the number of captured periods from some depth distances (pulses of specific dispersion values) which in turn limits their depth resolvability.


In this paper, a curved temporal LA is proposed in order to alleviate this problem and thus improve the depth resolution. By curving the LA, the more inclined rays corresponding to some periods of specific depth values find the chance to be recorded in the corresponding EIs as shown in Fig. \ref{curve1}. In this way, more number of  frequency cycles of the image of each depth would be recorded in the integral image array. This would make the corresponding frequency component of each depth in $\tilde{I}_s (\omega)$ narrower, improving its differentiability i.e., more depth resolution. There are various implementations of curved LA in spatial integral imaging systems. Among them, one promising scheme is the method in which instead of using a curved array, which has some practical difficulties, a large aperture lens is placed right in front of the lens array \cite{curvelargeaperture}. This large aperture lens provides a curving effect (Fig. \ref{curve1}). Using space-time analogy and the notion of temporal rays, we can adopt this technique to the temporal InI. As could be seen from Fig. \ref{curve1}, by using this method depth resolution has been improved due to capturing more frequency cycles from the image of each depth. In the next section, we prove quantitatively how this method increases depth resolution by numerical simulations.\\
\section{Results and analysis}\label{result}
\begin{figure*}[!ht]
  \centering
  \includegraphics[width=\linewidth]{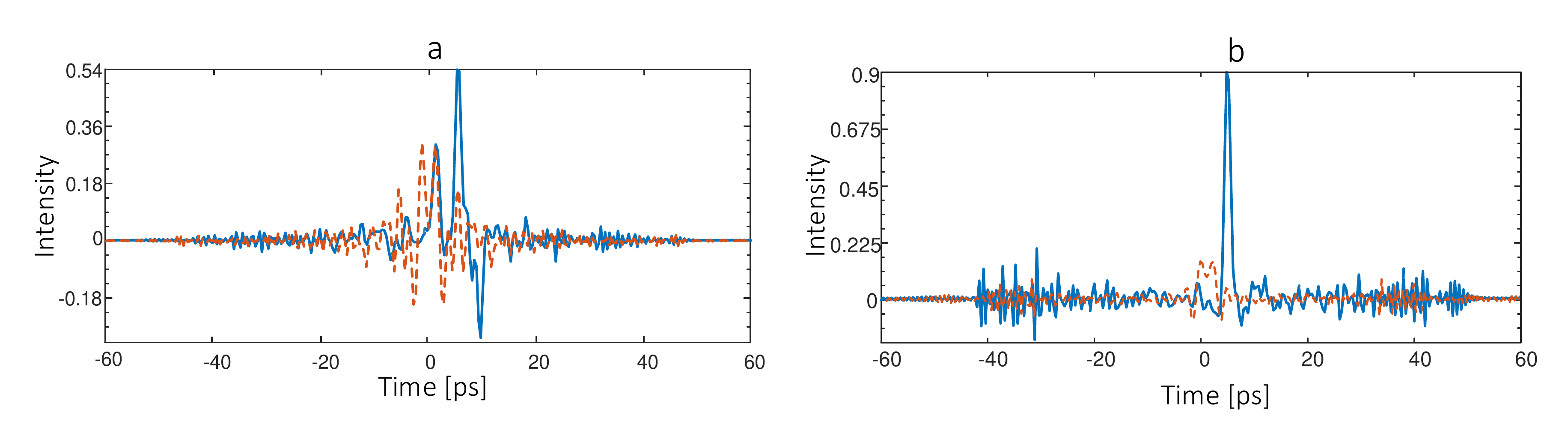}
  \caption{Reconstruction in the dispersive depth of the $I_1$ pulse (solid curve) with (a) conventional temporal InI and (b) temporal CInI imaging systems.}\label{rescurv}
\end{figure*}

In this section, we show numerical simulations of our proposed methods to confirm their superiority in view of DoF and depth resolution. First, we show the results of the temporal CInI system. The structure of the simulated temporal CInI, as discussed earlier, is the same as a flat temporal InI system except for a large aperture TL before the pick-up TL array and after the virtual display TL array. Both the TLs in the LA and the large aperture TLs have the same specifications in pick-up and display steps. Each of the TLs in the array has a temporal aperture equal to $10 Ps$ and their focal GDD is chosen in a way to set the temporal depth resolution of the system at $25 {ps}^2$. This means that the minimum resolvable GDD difference between pulses is $25 {ps}^2$. For the simulation of the temporal CInI, two Gaussian pulses of same temporal durations equal to $5 ps$ (full width at half maximum) are set as the input pulses with GDD values (from the pulse's initial position to the position of TL array) equal to 28.5 ${Ps}^2$ and 3.5 $ps^2$, respectively. We name the Gaussian pulse with the larger GDD value as $I_1$ and the other one which has a smaller GDD value as $I_2$. The large aperture TL has a temporal aperture duration of 50 $ps$. The curvature of this TL, which relates indirectly to its focal length, has been varied and accordingly, results have been generated for comparison. For simplicity, we name the large aperture TL as the curving TL (CTL).\\
 In this simulation, the number of TLs is 5. Figure \ref{rescurv} compares the reconstructed image of $I_1$ (solid pulse) in a temporal CInI and InI systems. For performing the curved lens array reconstruction, ray optics relations could not be used anymore. Therefore, the whole reconstruction set-up has been modeled based on wave-optics relations. For the flat lens-array this relation involves each elemental image with its corresponding TL transfer function, convolving each elemental image with the dispersion impulse response from the EI plane to the lens plane, passing through each lens transfer function, again convolving with the dispersion impulse response from the lens to the image plane and summation on all EI indexes:

\begin{eqnarray}\label{reconconv1}
  {O}^t(t,D) &=& \biggl(\sum_{a=0}^{m-1}\biggl\{\biggl[ \left(\sum_{{p}=0}^{m-1}{\tilde{I}_{{{p}}}(t)} \right )\otimes{\frac{e^\frac{it^2}{t\tilde{g}}}{\sqrt{2{\pi}i\tilde{g}}}}\biggr ]{\times}e^{\frac{-i\left(t-\tilde{s}_t{\times}\left(a-\frac{m-1}{2} \right ) \right )^2}{2D_f}} \\
   &&{\times}  {\Pi}{\left(\frac{t-\tilde{s}_t\times{\left(a-\frac{m-1}{2} \right )}}{\tilde{s}_t} \right )}\biggr\}\biggr){\otimes}{\frac{e^\frac{it^2}{2D}}{\sqrt{2{\pi}iD}}}. \nonumber
\end{eqnarray}

This reconstruction relation could be modified for the curved structure by adding the transfer function of the CTL after the transfer function of each TL as follows:

\begin{eqnarray}\label{reconconv2}
  {O}^t_c(t,D) &=& \biggl(\sum_{a=0}^{m-1}\biggl\{\biggl[ \left(\sum_{{p}=0}^{m-1}{{{}\tilde{I}_C}_{{p}}(t)} \right )\otimes{\frac{e^\frac{it^2}{t\tilde{g}}}{\sqrt{2{\pi}i\tilde{g}}}}\biggr ]{\times}e^{\frac{-i\left(t-\tilde{s}_t{\times}\left(a-\frac{m-1}{2} \right ) \right )^2}{2D_f}} \\
   && {\times}{\Pi}{\left(\frac{t-\tilde{s}_t\times{\left(a-\frac{m-1}{2} \right )}}{\tilde{s}_t} \right )}{\times}{e^{\frac{-i(t)^2}{2{D^{c}_{f}}}}\times{\Pi}(\frac{t}{T_c})}\biggr\}\biggr){\otimes}{\frac{e^\frac{it^2}{2D}}{\sqrt{2{\pi}iD}}}, \nonumber
\end{eqnarray}
where ${{}\tilde{I}_C}_{{p}}(t)$ is the $p$th elemental image captured by the curved TL array, $\Pi$ represents a rectangular function for temporal aperture, ${D^{c}_{f}}$ and $T_c$ are  the dispersive focal length and aperture size of the CTL, respectively.

\begin{figure*}[!ht]
  \centering
  \includegraphics[width=7cm]{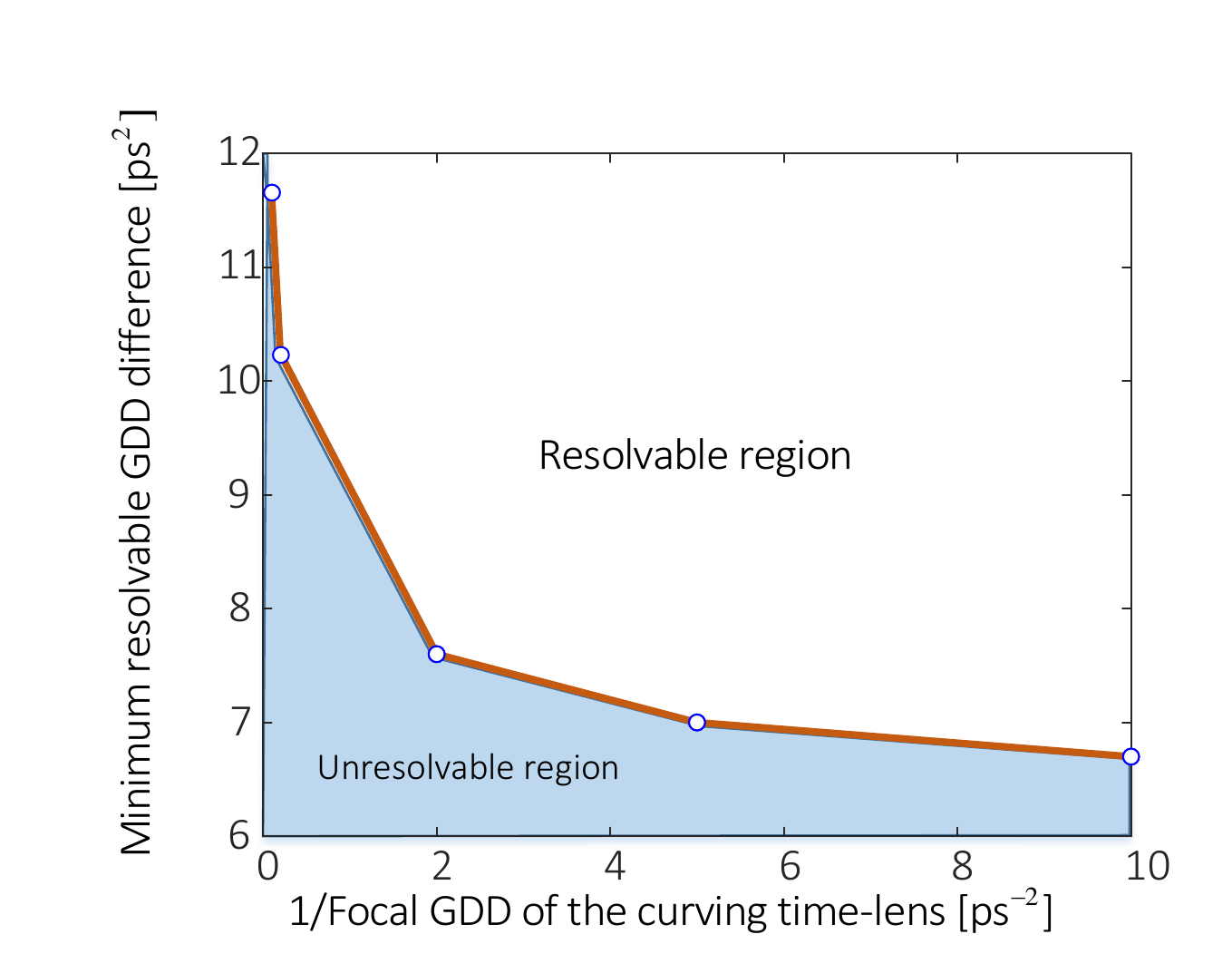}
  \caption{Resolvability enhancement via increasing the curvature of the curving lens.}\label{rescurv2}
\end{figure*}

As could be seen in Fig. \ref{rescurv}, in the case of conventional temporal InI structure, when GDD difference between input pulses is less than the depth resolution of the system (here, GDD difference is set as 10 $ps^2$) and reconstruction is performed at the dispersion value of $I_1$ pulse, its reconstructed image (blue curve in Fig. \ref{rescurv}(a)) becomes considerably distorted, as expected. Furthermore, the $I_2$ pulse also appears to some extent in reconstruction depth corresponding to the $I_1$ pulse which is undesirable. In contrast, in temporal CInI system, reconstruction at the depth distance corresponding to $I_1$ pulse (solid curve in Fig. \ref{rescurv} (b))leads to a less distorted image and also the intensity of the $I_2$ pulse, compared to conventional InI, becomes smaller at this reconstruction depth. In other words, this GDD difference is resolvable with CInI structure while it is smaller than the minimum resolvable depth distance of conventional InI. Thus, depth resolvability has been improved by using the curved LA structure. Assuming a focal GDD equal to 10 $ps^2$ for the CTL, depth resolvability improved from 25 $ps$ for conventional to 10 $ps$ for curved temporal InI system. In order to further evaluate and compare the performance of the CInI and conventional InI structures, the depth resolvability of the CInI system has been calculated for different curvature values of the CTL. In order to do this simulation, for each curvature value, the GDD of the $I_1$ pulse was assumed to be fixed at 28.5 $ps^2$ and the GDD of the $I_2$ pulse was increased from 3.2 $ps^2$ until the value where it is still resolvable from $I_1$. Fig. \ref{rescurv2} has been generated using a criterion we defined for the resolvability of reconstructed pulses. This criterion states that the reconstructed pulse could be resolved if its mean square error with the corresponding input pulse is less than 0.3. Fig. \ref{rescurv2} clearly shows that reducing value of the focal length of the CTL (increasing its curvature) improves the depth resolvability, as expected. \\
\begin{figure*}[!ht]
  \centering
  \includegraphics[width=\linewidth]{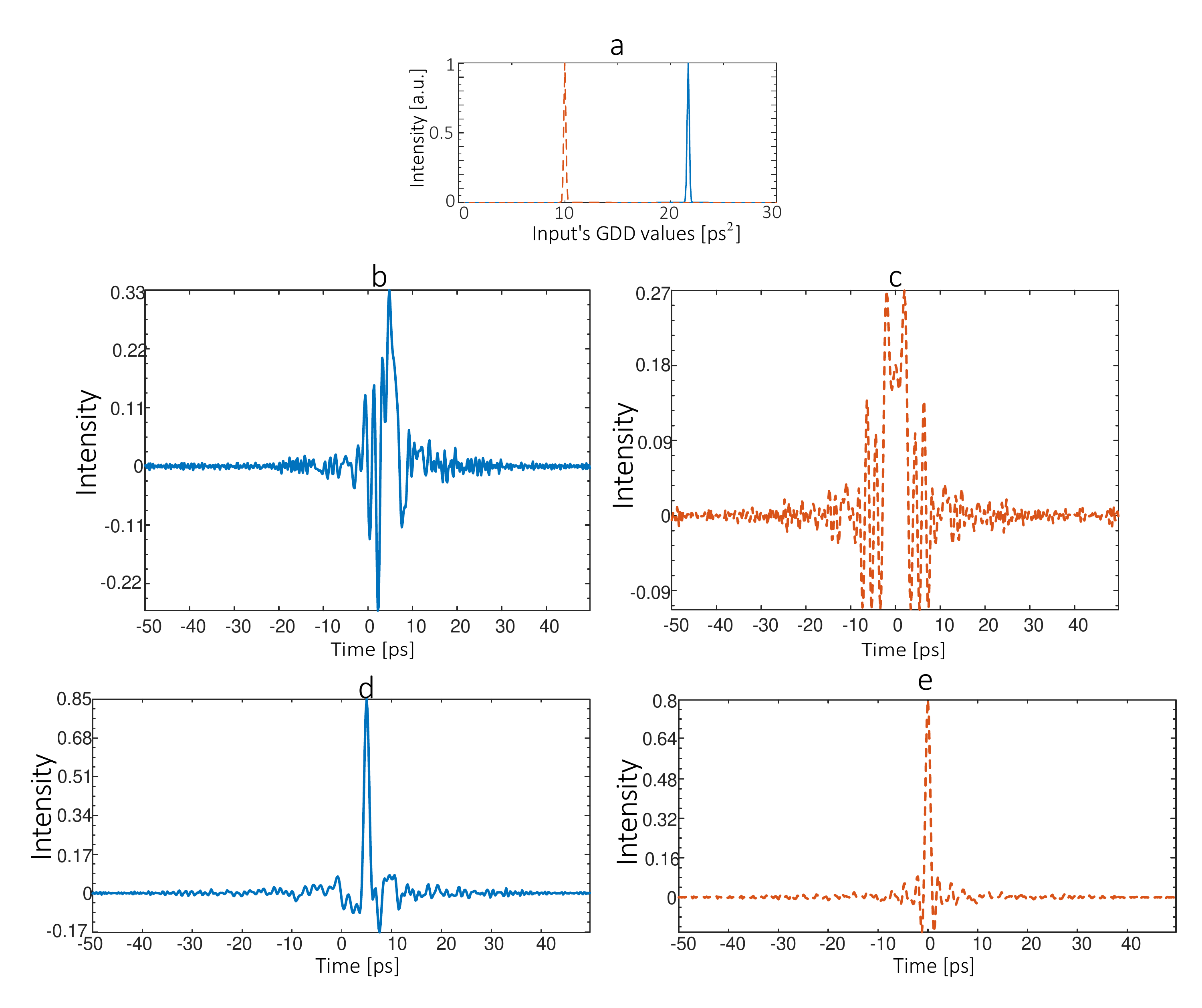}
  \caption{Simulation results for conventional and non-uniform temporal InI system. (a) input pulses with their corresponding GDD values. (b) and (c) reconstruction results for conventional system and (d) and (e) for non-uniform time-lens array.}\label{resnon}
\end{figure*}
\begin{figure*}[!hb]
  \centering
  \includegraphics[width=6cm]{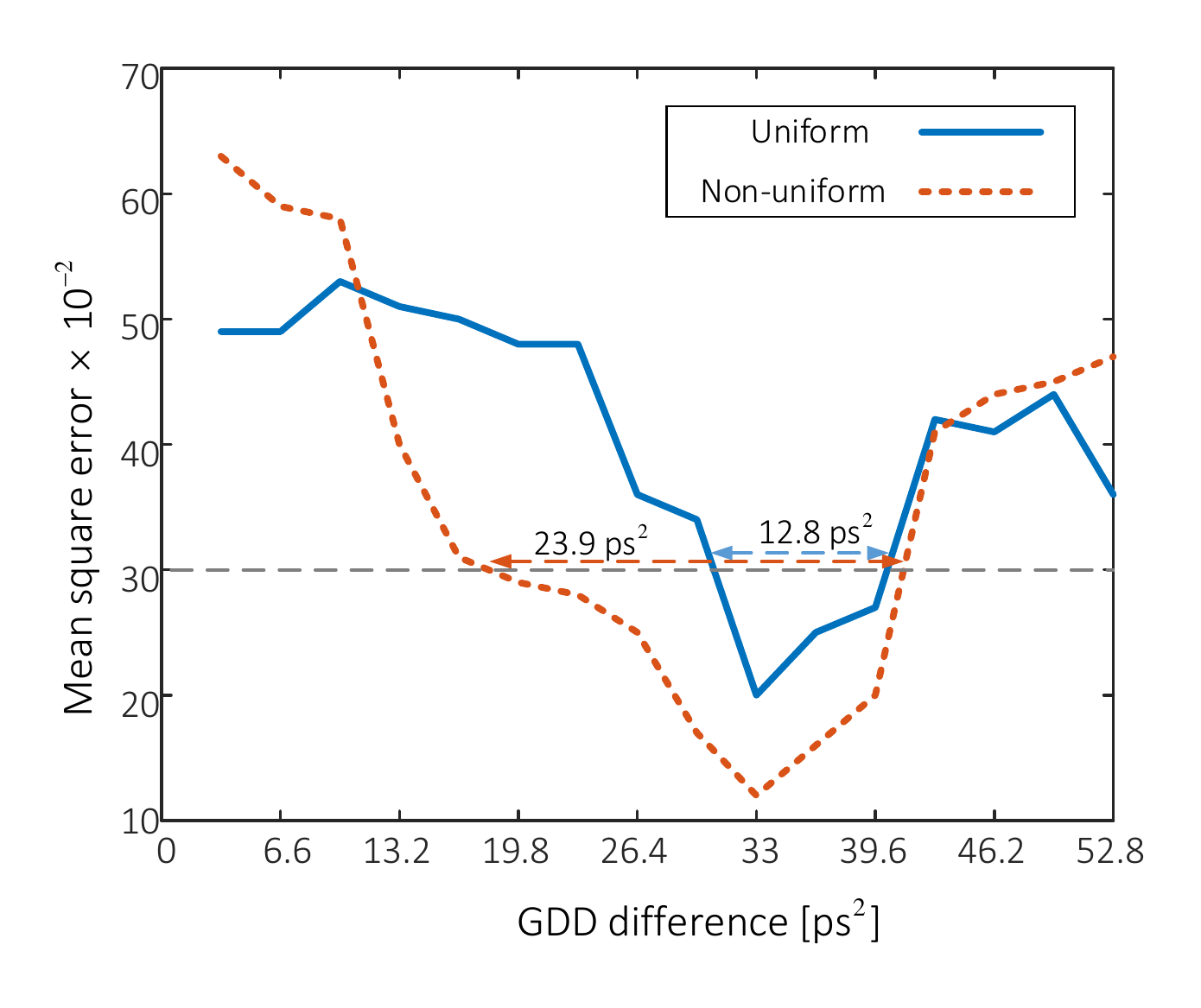}
  \caption{Comparison of DoF regions for conventional (uniform) and nonuniform TL array structures.}\label{resnon2}
\end{figure*}
In the second part of simulations, a numerical analysis was performed to show the DoF enhancement of the temporal InI system via the non-uniform TL array. The structure we chose for the TL array consists of $5$ TLs which two of them have focal GDDs equal to 1.25 $ps^2$ and the other $3$ TLs have focal GDDs equal to 1 $ps^2$. The input signal consists of two pulses of with temporal widths bigger than the lateral resolution limit of the system in order to only evaluate DoF parameter. These solid and dashed pulses have been placed in a considerably large dispersive distance from each other (larger than the DoF of the conventional temporal InI system) as shown in Fig. \ref{resnon} (a). According to the discussions in the previous section, we expect that in the conventional temporal InI system, the out-of-focus pulse, the pulse placed in the dispersive distance out of the DoF range, to become blurred (to lose its resolution). To perform the simulations and obtain results, we have reconstructed the image in the dispersive distance corresponding to the solid (Fig. \ref{resnon} (b)) and dashed (Fig. \ref{resnon} (c)) pulses for conventional uniform lens-array. The low resolution of the results is due to the reason that both pulses are placed out of the DoF of the conventional array from both sides. In contrast, the reconstruction results in the dispersive distance corresponding to the solid (Fig. \ref{resnon} (d)) and dashed (Fig. \ref{resnon} (e)) pulses for the non-uniform TL where the good resolution of the results shows the wide DoF of this structure which covers the dispersive distances of both pulses. Therefore, DoF region has been widened in the case of non-uniform TL array structure which is expected based on the presented discussions in the previous section.\\
  To demonstrate this improvement in DoF range more quantitatively, as shown in Fig. \ref{resnon2}, we have compared this region for conventional temporal InI and non-uniform temporal InI systems. To do this, the perfectness and resolution of the reconstruction results in different depth distances have been evaluated based on mean-square-error (MSE) of reconstructed pulse and input pulse. The value of MSE has been sketched versus the dispersive depth distance for conventional temporal InI and nonuniform temporal InI structures in Fig. \ref{resnon2}. As could be seen in this figure, the depth range with the acceptable lateral resolution has been increased in nonuniform structure from 12.8 $ps^2$ to 23.9 $ps^2$ meaning an almost twofold improvement in the DoF region.

\section{Conclusion}\label{con}
In this paper, first, we have analyzed a temporal counterpart of the spatial integral imaging technique. The depth information of a complex multi-dispersion pulse can be retrieved via this temporal integral imaging system. Two structures were proposed in order to improve depth-of-field (DoF) and depth resolution. In the first proposed structure, the depth resolvability has been improved through curving the time-lens array. The obtained improvement in depth resolvability has been analytically proven and also confirmed through numerical simulations. The second proposed structure uses a nonuniformity in the focal GDDs of time-lenses to increase the DoF region. The obtained improvement in the DoF range was approved via numerical simulations. Simulation results approved a wider DoF region with a factor of 1.87 through non-uniform time-lens array structure and 2.5 times improvement in depth resolvability via the proposed curved temporal InI structures.
\section*{Disclosures}

The authors declare no conflicts of interest.

\end{document}